\documentclass[conference]{IEEEtran}
\IEEEoverridecommandlockouts
\usepackage{cite}
\usepackage{amsmath,amssymb,amsfonts}
\usepackage{algorithmic}
\usepackage{graphicx}
\usepackage{textcomp}
\usepackage{xcolor}
\usepackage{flushend}
\def\BibTeX{{\rm B\kern-.05em{\sc i\kern-.025em b}\kern-.08em
    T\kern-.1667em\lower.7ex\hbox{E}\kern-.125emX}}

\newcommand{\mat}{\boldsymbol}
\newcommand{\SNR}{\textrm{SNR}}
\newcommand{\diag}{\textrm{diag}}
\newtheorem{lemma}{Lemma}

\IEEEoverridecommandlockouts
\IEEEpubid{\makebox[\columnwidth]{ 979-8-3503-6583-
2/25/\$31.00~\copyright~2025~IEEE \hfill}
\hspace{\columnsep}\makebox[\columnwidth]{ }}

\begin{document}

\title{URLLC Networks enabled by STAR-RIS,\\ Rate Splitting, and Multiple Antennas \\
\thanks{This work is supported by the European Commission’s Horizon Europe, Smart Networks and Services Joint Undertaking program under grant agreement 101139282, 6G-SENSES project. The work of Eduard A. Jorswieck was supported in part by the Federal Ministry of Education and Research (BMBF, Germany) through the Program of “Souver\"an. Digital. Vernetzt.” Joint Project 6G-Research and Innovation Cluster (RIC) under Grant 16KISK031.}
}

\author{\IEEEauthorblockN{Eduard A. Jorswieck$^1$, Mohammad Soleymani$^2$, Ignacio Santamaria$^3$, Jesús Gutiérrez$^4$}
\vspace{0.1cm}
\IEEEauthorblockA{
$^1$ Institute for Communications Technology, TU Braunschweig, Germany \\
$^2$ Signal and System Theory Group, University of Paderborn, Germany
 \\
$^3$ Dept. of Communications Engineering, University of Cantabria, Spain \\
$^4$ IHP - Leibniz-Institut für Innovative Mikroelektronik, Frankfurt (Oder), Germany}
}

\maketitle

\begin{abstract}
The challenges in dense ultra-reliable low-latency communication networks to deliver the required service to multiple devices are addressed by three main technologies: multiple antennas at the base station (MISO), rate splitting multiple access (RSMA) with private and common message encoding, and simultaneously transmitting and reflecting reconfigurable intelligent surfaces (STAR-RIS). Careful resource allocation, encompassing beamforming and RIS optimization, is required to exploit the synergy between the three. We propose an alternating optimization-based algorithm, relying on minorization-maximization. Numerical results show that the achievable second-order max-min rates of the proposed scheme outperform the baselines significantly. MISO, RSMA, and STAR-RIS all contribute to enabling ultra-reliable low-latency communication (URLLC). 
\end{abstract}

\begin{IEEEkeywords}
Wireless Networks, Interference Management, URLLC, BD-RIS, RSMA, MAC, optimization, resource allocation.
\end{IEEEkeywords}

\section{Introduction}

The requirements for the sixth generation of mobile communications (6G) will increase dramatically compared to 4G and 5G. On the one hand, the number of wireless devices and their quality of service requirements in terms of data rate, reliability and latency are growing. On the other hand, the capabilities of network nodes in terms of computing power and storage are also improving \cite{Tataria2021}. While there is some balance between demand and capability, one important resource that is not increasing is spectrum and available bandwidth. Therefore, moving to higher frequency bands (mmWave and THz) is considered to provide more available bandwidth, but will also require investment in new infrastructure and hardware as well as smart spatial processing. Next generation multiple access techniques \cite{Jorswieck24} are key to master the upcoming challenges. 

\subsection{State of the Art and Motivation} 

There exists several technologies considered for 6G, which are enabling specialized interference management and spatial processing to enhance the spectral efficiency, reliability, and low latency \cite{CIT-129}: 
\begin{itemize}
	\item Reconfigurable intelligent surfaces (RIS) allow to modify the radio propagation environment by choosing reflection coefficients. They are applied to boost signal power and reduce interference leakage \cite{DiRenzo2019}. Their capabilities depend on the hardware realization and variants such as simultaneously transmitting and reflecting (STAR) RIS \cite{9774942} and beyond diagonal (BD) RIS are studied \cite{li2022beyond}. 
	\item Multiple antenna systems (MIMO) including its massive variant (mMIMO) allow to separate devices in the spatial domain by transmit and receive beamforming \cite{bjornson2014massive}. The corresponding multiple access scheme is spatial division multiple access (SDMA). 
	\item Rate splitting multiple access (RSMA) is a coding and decoding technique which supports partial interference cancellation and includes both SDMA and non-orthogonal multiple access (NOMA) as special cases \cite{Clerckx23}. 
\end{itemize}
Since we target ultra-reliable low-latency communication (URLLC), the finite block length (FBL) achievable rates are considered \cite{polyanskiy2010channel}. The technologies mentioned above are well researched, and there exists a large body of work on their combinations and performance optimization. The following comparison includes some selected recent references. While \cite{nasir2020resource} and \cite{ghanem2020resource} study resource optimization for multi-user FBL MISO systems, RIS are not considered. The papers in \cite{soleymani2022noma,pan2020multicell,soleymani2023noma} optimize resource allocation and RIS coefficients for multiuser multiple-input single-output (MISO) or MIMO systems, but consider first-order rates. In \cite{vu2022intelligent} and \cite{almekhlafi2021joint} both RIS, and multiuser FBL, but not multiple antennas, are considered. All of the mentioned works, do not include RSMA. In \cite{xu2022rate,ou2022resource}, a network with RSMA in combination with FBL, but without RIS is optimized. In \cite{soleymani2022rate, soleymani2023energy, soleymani2023rate}, RSMA schemes are developed for MIMO systems, considering the first-order rate. Finally, in \cite{dhok2022rate}, a multiuser network with RSMA, FBL, and STAR RIS is optimized. However, the spatial dimension with multiple antennas is missing.

The motivation of the current paper is to close the gap shown in the state of the art section above. We consider a single-cell multiuser MISO network which is supported by STAR RIS, rate splitting, and applies FBL in the computation of user rates. The optimization problem considered in this paper is the maximization of worst-case user rates. Spatial precoding and RIS coefficients are the optimization variables, while there are transmit power and RIS coefficient constraints. The proposed alternating optimization (AO) solution is low complex, but provides significant gains in the setup with both RSMA and STAR-RIS. Our results show that the careful exploitation of the spatial and code domain enables URLLC links. 

\subsection{Organization and Outline}

In the next section (Section \ref{sec:smps}), we provide the system model and develop the optimization problem considered in the paper including objective, constraints, and their operational meaning in the particular network setup. Then in Section \ref{sec:aora}, we present the AO-based resource allocation algorithm, present its properties and discuss its limitations. In Section \ref{sec:ni}, we present numerical simulation results to compare the achieved performance to baselines. The paper is concluded in Section \ref{sec:conc}.

\section{System Model and Problem Statement}
\label{sec:smps}

\begin{figure}[ht]
	\centering
	\includegraphics[width=.8\linewidth]{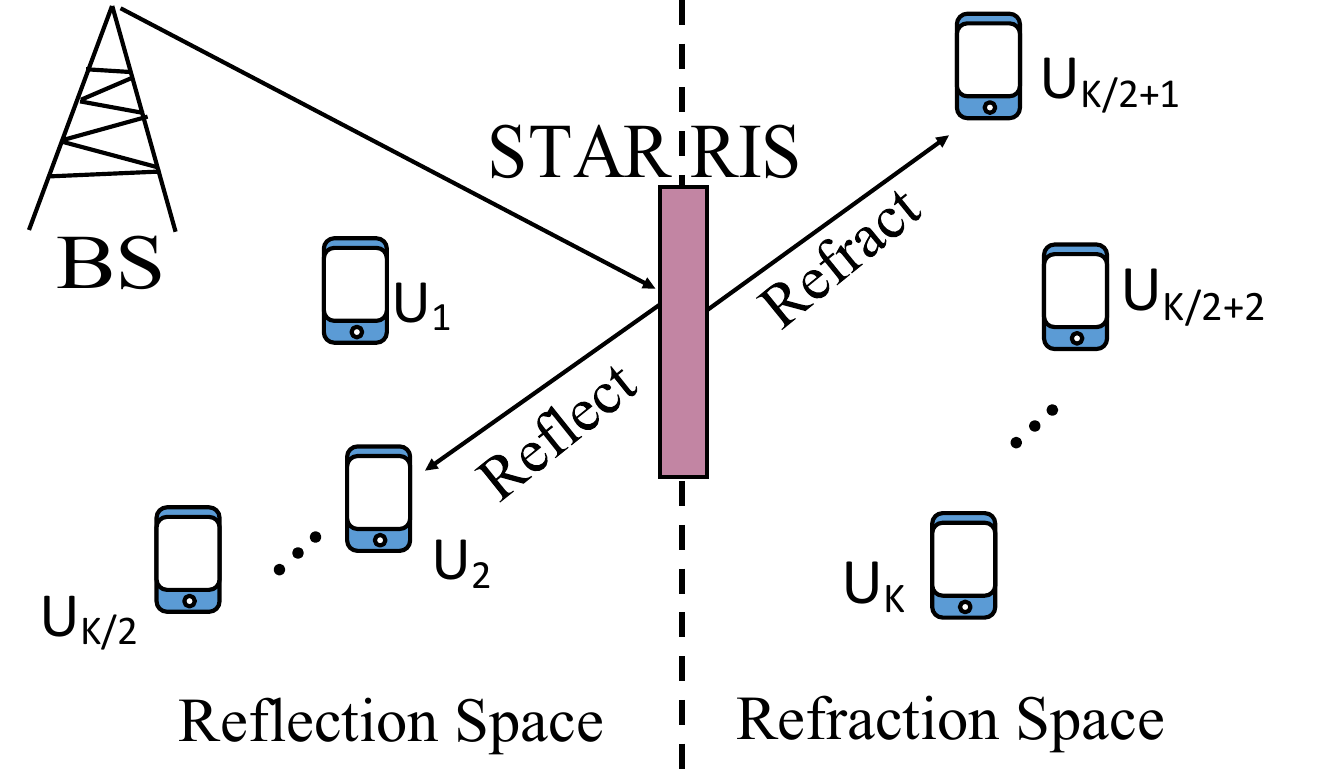}
	\caption{Network model: STAR RIS assisted downlink transmission. }
	\label{fig:network-model}
\end{figure}

We consider a single-cell setup as shown in Fig. \ref{fig:network-model}. A multiple antenna base station (BS) serves $K$ single-antenna users (MISO broadcast channel). The STAR RIS is located centrally and serves one half of the users in the reflect side ($u_1 ,\cdots,u_{K/2}$) and one half of the users in the transmit side ($u_{K/2+1},\cdots,u_K$). 

\subsection{Channel Model}

We assume that the channel between the BS and user $k$ is given by 
\begin{eqnarray}
	\label{eq:cm}
	\mat{h}_k = \mat{f}_k \mat{\Theta}^{r/t} \mat{G} + \mat{d}_k, 
\end{eqnarray}
where $\mat{d}_k$ is the direct channel between the BS and user $k$, while the product of the three terms is the in-direct channel over the RIS. $\mat{G}$ is the MIMO channel from the BS to the RIS, $\mat{f}_k$ is the channel from the RIS to the user $k$. The RIS coefficients are collected in the matrix $\mat{\Theta}^{r/t}$. For users $1,\cdots,K/2$, we use $\mat{\Theta}^r$ and for users $K/2+1,\cdots,K$ we apply $\mat{\Theta}^t$. 

The RIS constraints for the reflect and transmit\footnote{We use transmit space and refraction space alternately.} side, are modeled as standard diagonal elements $\diag(\theta_1,\cdots,\theta_M)$, where each pair of elements have amplitudes sum up to one \cite{liu2021star}, i.e. 
\begin{eqnarray}
	|\theta_m^t|^2 + |\theta_m^r|^2 = 1 \label{eq:thecon}.  
\end{eqnarray}
The phases of reflective and refractive parts of the RIS elements can be optimized separately. In this work, we consider mode switching (MS), i.e., 
\begin{eqnarray}
    |\theta_m^t|^2 & = & 1, |\theta_m^r|^2 =0, \quad k \in \{1,\cdots,M/2\} \nonumber \\
    |\theta_m^t|^2 & = & 0, |\theta_m^r|^2 = 1, \quad k \in \{M/2+1,\cdots,M\} \nonumber. 
\end{eqnarray}
A more relaxed constraint set is obtained by replacing the equalities by inequalities (smaller than or equal). In the numerical assessments, we consider both sets, and denote the strict equality set with an index $I$. 

\subsection{Signal Model}

We consider a 1-layer rate splitting model, where one common data stream is transmitted together with $K$ private data streams. 
The signal sent from the BS is a combination of the $K$ codewords corresponding to the private messages of the users and the common codeword
\begin{eqnarray}
	\mat{x} = \mat{w}_c d_c + \sum_{k=1}^K \mat{w}_k d_k \label{eq:txsig}, 
\end{eqnarray}
where the common data codeword is denoted by $d_c$ and the private data codewords by $d_1,\cdots,d_K$. The beamforming vectors for the common codeword are $\mat{w}_c$ and for the private codewords $\mat{w}_1,\cdots,\mat{w}_K$, respectively. The symbols of the codewords of the common and the private messages are statistically independent. 

The received signal of user $k$ is given by 
\begin{eqnarray}\label{eq:recsig}
	y_k = \mat{h}_k \mat{x} + n_k, 
\end{eqnarray}
where $n_k \sim \mathcal{CN}(0,\sigma^2)$ is additive white Gaussian noise with noise variance $\sigma^2$. Inserting (\ref{eq:txsig}) into (\ref{eq:recsig}) gives 
\begin{eqnarray}
	y_k = \mat{h}_k \mat{w}_c d_c + \mat{h}_k \mat{w}_k d_K + \sum_{l \neq k} \mat{h}_k \mat{w}_l d_l + n_k \label{eq:recsig2}. 
\end{eqnarray}
The first term in (\ref{eq:recsig2}) corresponds to the desired common signal, the second term to the desired private signal, while the third term corresponds to multiple access interference (MAI) from the signals intended for the other users.

\subsection{Achievable Rate Expressions}

Every user first decodes the desired common message, subtracts it from the received signal, and then continues with its own desired message. The private signals of the other users are treated as noise (TIN). The second-order rate achievable for the common message at user $k$ is given by
\begin{eqnarray}
	r_{c,k} = \log(1 + \SNR_{c,k}) - Q^{-1}(\epsilon_c) \sqrt{ \frac{V_{c,k}}{n_c}}, \label{eq:cr}
\end{eqnarray}
with signal-to-noise-ratio $\SNR_{c,k}$ defined below, inverse Q-function $Q^{-1}$, error probability for the common message $\epsilon_c$, block length $n_c$ for the codewords belonging to the common message, and dispersion $V_{c,k}$. The SNR for the common message is computed as 
\begin{eqnarray}
	\SNR_{c,k} = \frac{ | \mat{h}_k \mat{w}_c|^2 }{\sigma^2 + \sum_{l=1}^K |\mat{h}_l \mat{w}_l|^2 }. 
\end{eqnarray}
The channel dispersion term for Gaussian signals and Gaussian interference can be computed as \cite{scarlett2016dispersion}
\begin{eqnarray}
	V_{c,k} = 2 \frac{ \SNR_{c,k}}{1 + \SNR_{c,k}} \label{eq:disp}.
\end{eqnarray}
The common message should be decodable (with error probability $\epsilon_c$) at all receivers. Therefore, its rate must fulfill 
\begin{eqnarray}
	r_c \leq \min_k r_{c,k} \nonumber. 
\end{eqnarray}
After subtracting the common message, the $k$-th receiver decodes it desired private message. The second-order rate is given by
\begin{eqnarray}
	r_{p,k} = \log( 1 + \SNR_{p,k}) - Q^{-1}(\epsilon_k)\sqrt{ \frac{V_{p,k}}{n_k}}, \label{eq:pr}
\end{eqnarray}
where $\epsilon_k$ is the error probability of the private message of user $k$, $n_k$ is the codeword length of user $k$, and the SNR for the private message of user $k$ is computed as 
\begin{eqnarray}
	\SNR_{p,k}= \frac{ |\mat{h}_k \mat{w}_k|^2}{\sigma^2 + \sum_{l=1,l \neq k}^K |\mat{h}_l \mat{w}_l|^2 }. 
\end{eqnarray}
The final achievable second-order rate of user $k$ is given by 
\begin{eqnarray}
	R_k = r_c + r_{p,k}
	\nonumber.
\end{eqnarray}

\subsection{Problem Statement}

We consider the following optimization problem 
\begin{eqnarray}
	& \max\limits_{\mat{w}_c,\mat{w}_1,\cdots,\mat{w}_K, \mat{\Theta}^r, \mat{\Theta}^t} & \qquad \min_{k \in \{1,\cdots,K\}} R_k \label{eq:opt} \\
	& \textrm{s.t.} & ||\mat{w}_c||^2 + \sum_{k=1}^K ||\mat{w}_k||^2 \leq P \label{eq:con1} \\
	& & \{\mat{\Theta}^t, \mat{\Theta}^r\} \in \mathcal{T} \label{eq:con2}, 
\end{eqnarray}
where the objective function in (\ref{eq:opt}) is the max-min second-order achievable rate, the first constraint in (\ref{eq:con1}) is the transmit power constraint and the second constraint in (\ref{eq:con2}) is the RIS constraint. 

Note that the optimization problem in (\ref{eq:opt}) contains implicitly power optimization and thereby rate optimization for the common and private messages. This can be seen if the beamforming vectors are decomposed into unit-norm vectors and power allocations, i.e., 
\begin{equation}
	\mat{w}_k = \sqrt{p_k} \mat{u}_k, \nonumber
\end{equation}
with power allocation $p_k$ and unit norm beamforming vector $||\mat{u}_k||=1$. 

\section{AO-based Resource Allocation and Optimization}
\label{sec:aora}

In \cite{Soleymani24}, a framework for problems with a structure similar to (\ref{eq:opt}) is developed. The approach is based on AO, i.e., we fix the RIS coefficients $\mat{\Theta}^t$ and $\mat{\Theta}^r$ and optimize the beamforming vectors $\mat{w}_c$, $\mat{w}_1,\cdots,\mat{w}_K$. Next, for fixed beamforming vectors, the two RIS coefficient matrices $\mat{\Theta}^t$ and $\mat{\Theta}^r$ are optimized. These two steps are repeated until convergence, i.e., until the utility function does not improve by more than $\varepsilon>0$. 

\subsection{Beamforming Optimization}

It is well known that the max-min fairness design for MISO broadcast  channels is a difficult programming problem \cite{Naghsh2019}. It is usually approached by the minorization-maximization technique. Therefore, we follow a similar approach. First, we compute suitable concave lower bounds for the rate expressions satisfying the conditions in \cite[Section II.B]{aubry2018new}. 

\begin{lemma}
	\label{lem:1}
	Concave lower bounds for the rate expressions at the points $\mat{w}_k^t, \mat{w}_c^t$ are given by 
	\begin{eqnarray}
		r_{c,k} & \geq & \tilde{r}_{c,k} = a_{c,k} + \frac{ 2 \Re{ \{ (\mat{h}_k \mat{w}_c^t)^* \mat{h}_k \mat{w}_c \} }}{\sigma^2 + \sum_{l=1}^K |\mat{h}_k \mat{w}_l^t|^2 + |\mat{h}_k \mat{w}_c^t} \nonumber \\ & + & \frac{2 Q^{-1}(\epsilon_c)}{\sqrt{n_c V_{c,k}^t}} \frac{ \sigma^2 + \sum_{k=1}^K \Re{ \{(\mat{h}_k \mat{w}_c^t)^* \mat{h}_k \mat{w}_c \}}}{\sigma^2 + \sum_{l=1}^K |\mat{h}_k \mat{w}_l^t|^2 + |\mat{h}_k \mat{w}_c^t|^2} \nonumber \\ 
		& - & b_{c,k} \frac{ \sigma^2 + \sum_{l=1}^K |\mat{h}_k \mat{w}_l|^2 + |\mat{h}_k \mat{w}_c|^2}{\sigma^2 + \sum_{l=1}^K |\mat{h}_k \mat{w}_l^t|^2 + |\mat{h}_k \mat{w}_c^t|^2}, \label{eq:inq}
	\end{eqnarray}
	where $t+1$ is the current iteration, $a_{c,k}$ and $b_{c,k}$ are constants and given by
	\begin{eqnarray}
		a_{c,k} & = & \log( 1 + \SNR_{c,k}^t) - \SNR_{c,k}^t \nonumber \\ & - & \frac{Q^{-1}(\epsilon_c)}{\sqrt{n_c}} (\sqrt{V_{c,k}^t}/2 + 1/\sqrt{V_{c,k}^t}) \label{eq:ack} \\
		b_{c,k} & = & \SNR_{c,k} + \frac{\zeta_{k,c} Q^{-1}(\epsilon_c)}{\sqrt{n_c + V_{c,k}^t}} \label{eq:bck}.
	\end{eqnarray}
	The $\SNR_{c,k}^t$ and $V_{c,k}^t$ are obtained by computing the SNR and the dispersion at the points $\mat{w}_k^t, \mat{w}_c^t$. Finally, 
	\begin{eqnarray}
		\zeta_{k,c} = \frac{ \sigma^2 + \sum_{l=1}^K |\mat{h}_k \mat{w}_l^t|^2}{\sigma^2 + \sum_{l=1}^K |\mat{h}_k \mat{w}_l^t|^2 + |\mat{h}_k \mat{w}_c^t|^2} \nonumber. 
	\end{eqnarray}
	The concave lower bound for the rate $r_{p,k}$ looks very similar, except that the interference term from the private message is removed. 
\end{lemma}

The proof follows similar steps as in the proofs of \cite[Lemma 3]{soleymani2023spectral} and \cite[Lemma 1]{Soleymani24} and is omitted.

In the optimization problem (\ref{eq:opt}), we replace the rate expressions for $r_{c,k}$ and $r_{p,k}$ with their concave lower bounds $\tilde{r}_{c,k}$ and $\tilde{r}_{p,k}$. The resulting problem is convex and can be solved by standard convex solvers \cite{boyd2004convex}. 

\subsection{RIS Optimization}

For fixed beamforming vectors $\mat{w}_c$ and $\mat{w}_1,\cdots, \mat{w}_K$ obtained in the last subsection, we now optimize the RIS coefficient matrices $\mat{\Theta}^r$ and $\mat{\Theta}^t$. Since the channels $\mat{h}_k$ (as a function of the RIS matrices) and the beamforming vectors are symmetric in the rate expressions, it is possible to re-use the approach from Lemma \ref{lem:1} for the channel vectors. As shown in (\ref{eq:cm}), the channel vectors depend linearly on the RIS matrices $\mat{\Theta}^{r/t}$. This approach is detailed in \cite[Corollary 1]{Soleymani24} and is not repeated here. 

After replacing the rate expressions with their concave lower bounds, we still have the challenge with the non-convex constraints for MS STAR RIS matrices. The constraints for mode switching can be expressed as
\begin{eqnarray}
	|\theta_m^t|^2 \leq 1 \label{eq:thecon1} \\
	|\theta_m^t|^2 \geq 1 \label{eq:thecon2}, 
\end{eqnarray}
for $m=1,\cdots,M/2$ and analogue for $\theta_m^r$ for $m=M/2+1,,\cdots,M$.
The second constraint in (\ref{eq:thecon2}) is convexified by the concave convex procedure (CCP) to arrive at 
\begin{eqnarray}
	& & |\theta_m^r(t)|^2 + 2 \Re \{ \theta_m^r(t) (\theta_m^r - \theta_m^r(t))^*\} \nonumber \\ & & + |\theta_m^t(t)|^2 + 2 \Re \{ \theta_m^t(t) (\theta_m^t - \theta_m^t(t))^*\} \geq (1-\varepsilon). \nonumber
\end{eqnarray}
The convexified programming problem is solved and then the solution is projected to the feasible set. 

\section{Numerical Illustrations}
\label{sec:ni}

In the simulation scenario, half of the users are in the reflection space, while the other half are in the refraction space, compare to Fig. \ref{fig:network-model}. Hence, a reflective RIS cannot assist half of the users. Moreover, we consider only a mode switching scheme for the STAR-RIS.  

There are a total of six users, three of them in the reflection space and three in the refraction space. The BS has four antennas. The STAR-RIS has in total 24 elements. The error probability is set to $\epsilon_c=\epsilon_k= 10^{-5}$ for all $k=1,\cdots,K$. The block length is set to $n_c=n_k=256$ for all $k=1,\cdots,K$. 

We assume that there is a line-of-sight (LOS) link between the BS and the RIS as well as between the RIS and each user from either reflect or refract side. Hence, the small-scale fading of $\mat{G}$ and $\mat{f}_k$ in (\ref{eq:cm}) is described by Ricean fading according to \cite[(55)]{pan2020multicell} with a Ricean factor of three. For the direct link, we model it as non-LOS with small-scale Rayleigh fading. The large scale fading is modeled according to the path-loss model in \cite[(59)]{soleymani2022improper}. The other propagation parameters, including the antenna gains, bandwidth, noise power density, path loss components, and the path loss at the reference distance of $1$ meter, are based on \cite{soleymani2022improper}. 

\begin{figure}[ht]
    \centering
    \includegraphics[width=\linewidth]{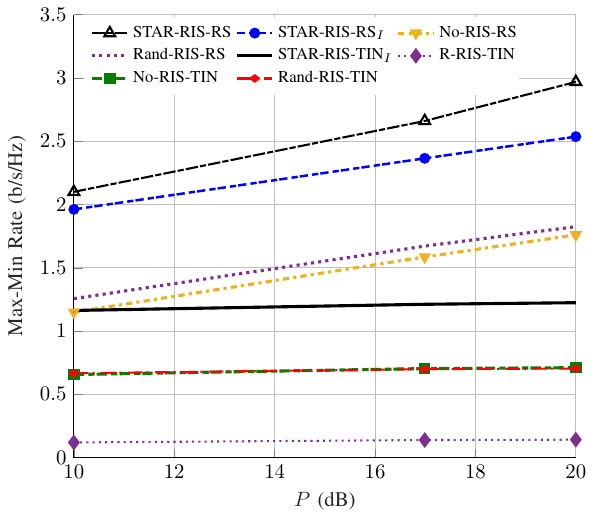}
    \caption{Max-min rate over transmit power.}
    \label{fig:sim1}
\end{figure}

In Fig. \ref{fig:sim1}, we compare the max-min second-order rates in [bits/s/Hz] over the transmit power in [dB] for the following schemes: 
the first baseline is \textbf{R-RIS-TIN}, where only reflective RIS coefficients are chosen, and all interference is treated as noise (TIN, no common message). In this case the system is interference limited and increasing transmit power does not improve max-min rates. The second baseline is \textbf{No-RIS-TIN}, where no RIS is applied and TIN is performed. Interestingly, the performance without RIS is similar to random RIS elements. Still the system is interference limited. 

The first scheme with rate splitting is \textbf{No-RIS-RS}, where no RIS is applied, but RSMA. The system is noise limited, because RSMA can handle the interference better than multiple antennas at the BS alone. The random RIS with RSMA improve a little bit performance (\textbf{Rand-RIS-RS$_{I}$}). The second best performance is obtained by \textbf{STAR-RIS-RS$_{I}$} where STAR-RIS is combined with RSMA for MS with equality constraints. The best performance is obtained by \textbf{STAR-RIS-RS}, where the equality constraint (\ref{eq:thecon}) is relaxed to smaller than or equal. 

There are several important observations from the Fig. \ref{fig:sim1}: 
\begin{itemize}
	\item Without RSMA the system is interference limited since four antennas cannot remove interference between six receivers. 
	\item RSMA brings the system to max-min rates which scale with $P$, the system becomes more noise-limited. 
	\item The gain obtained by STAR-RIS in combination with RSMA is significant. 
\end{itemize}
	
\section{Conclusions and Future Work}
\label{sec:conc}

We have studied the synergies of multiple antenna systems, with STAR-RIS, and RSMA. All three technologies contribute to enable URLLC in the downlink. We have formulated the max-min second-order rate problem and provided an iterative algorithm to find a low-complexity solution. The numerical simulations indicate that all three technologies improve the achievable performance compared to the baseline schemes where only one or two technologies are used. 

In future work, we plan to extend the system model to include further performance targets, including secrecy and sensing. Furthermore, the channel model will be extended towards higher frequency bands, e.g., mmWave and THz, to understand the synergies of the technologies for these spectrum bands, too. 

\bibliographystyle{ieeetr}
\bibliography{ref2.bib}

\end{document}